\documentclass[twocolumn,twocolappendix]{aastex63}
\usepackage{multirow}

\newcommand{\kms}{\mbox{km s$^{-1}$}}                                    

\shorttitle{ALMA Band 7 Survey of PMC Disks}
\shortauthors{Wu et al.}
  
\begin{document}
\title{ALMA 0.88 mm Survey of Disks around Planetary-mass Companions}


\author{Ya-Lin Wu}
\altaffiliation{51 Pegasi b Fellow; yalinwu@utexas.edu}
\affiliation{McDonald Observatory and the Department of Astronomy, University of Texas at Austin, Austin, TX 78712, USA}

\author{Brendan P. Bowler}
\affiliation{McDonald Observatory and the Department of Astronomy, University of Texas at Austin, Austin, TX 78712, USA}

\author{Patrick D. Sheehan}
\altaffiliation{CIERA Fellow}
\affiliation{Center for Interdisciplinary Exploration and Research in Astrophysics and Department of Physics and Astronomy,
Northwestern University, 2145 Sheridan Road, Evanston, IL 60208-3112, USA}

\author{Sean M. Andrews}
\affiliation{Harvard-Smithsonian Center for Astrophysics, 60 Garden Street, Cambridge, MA 02138, USA}

\author{Gregory J. Herczeg}
\affiliation{Kavli Institute for Astronomy and Astrophysics, Peking University, Beijing 100871, People's Republic of China}

\author{Adam L. Kraus}
\affiliation{McDonald Observatory and the Department of Astronomy, University of Texas at Austin, Austin, TX 78712, USA}

\author{Luca Ricci}
\affiliation{Department of Physics and Astronomy, California State University Northridge, 18111 Nordhoff Street, Northridge, CA 91330, USA}

\author{David J. Wilner}
\affiliation{Harvard-Smithsonian Center for Astrophysics, 60 Garden Street, Cambridge, MA 02138, USA}

\author{Zhaohuan Zhu}
\affiliation{Department of Physics and Astronomy, University of Nevada, Las Vegas, 4505 S. Maryland Pkwy, Las Vegas, NV 89154, USA \\ \\ {\normalsize Accepted for publication in AJ}}

\begin{abstract} 
Characterizing the physical properties and compositions of circumplanetary disks can provide important insights into the formation of giant planets and satellites. We report ALMA 0.88 mm (Band 7) continuum observations of six planetary-mass (10--20 $M_{\rm Jup}$) companions: CT Cha b, 1RXS 1609 b, ROXs 12 b, ROXs 42B b, DH Tau b, and FU Tau b. No continuum sources are detected at the locations of the companions down to 3$\sigma$ limits of 120--210 $\mu$Jy. Given these non-detections, it is not clear whether disks around planetary-mass companions indeed follow the disk flux-host mass trend in the stellar regime. The faint radio brightness of these companion disks may result from a combination of fast radial drift and a lack of dust traps. Alternatively, as disks in binary systems are known to have significantly lower millimeter fluxes due to tidal interactions, these companion disks may instead follow the relationship of moderate-separation binary stars. This scenario can be tested with sensitive continuum imaging at rms levels of $\lesssim$10~$\mu$Jy.
\end{abstract}


\section*{\textbf {\normalsize1. I\lowercase{ntroduction}}}

\begin{deluxetable*}{@{}lccccllc@{}}
\tablewidth{\linewidth}
\tablecaption{ALMA 880~\micron~Survey of Planetary-mass Companions \label{tb:obs}}
\tablehead{
\colhead{Source} &
\colhead{Date} &
\colhead{$N_{\rm ant}$} &
\colhead{$L_{\rm baseline}$} &
\colhead{$T_{\rm int}$\tablenotemark{$\dagger$}} &
\colhead{Calibrators} & 
\colhead{Beam} & 
\colhead{rms}
	\\
	\colhead{ }  &	
	\colhead{ }  &
	\colhead{ } &
	\colhead{(m)} &	
	\colhead{(s)} &
	\colhead{Gain/Bandpass/Flux} &
	\colhead{(size, PA)} &
	\colhead{($\mu$Jy beam$^{-1}$)} 
}
\startdata 
CT Cha		& 2016-12-17	& 42 & 15.1--460.0  & 1609	& J1058$-$8003, J1427$-$4206, J1107$-$4449 &	0\farcs87$\times$0\farcs58, 6\fdg4	& 52	\\
1RXS 1609	& 2017-01-29	& 43	& 15.1--331.0  & 941		& J1634$-$2058, J1427$-$4206, Titan &	0\farcs97$\times$0\farcs70, 72\fdg9	& 45	\\
ROXs 12		& 2017-04-08	& 41	& 15.1--390.0  & 729		& J1625$-$2527, J1517$-$2422, J1517$-$2422 &	1\farcs22$\times$0\farcs76, 99\fdg7	& 71	\\
ROXs 42B 	& 2017-04-13	& 44 & 15.1--460.0  & 941		& J1625$-$2527, J1517$-$2422, J1517$-$2422 &	0\farcs69$\times$0\farcs57, 87\fdg8	& 43	\\
DH Tau 		& 2018-05-13	& 45	& 15.1--313.7  & 1730	& J0438$+$3004, J0510$+$1800, J0510$+$1800 &	1\farcs17$\times$0\farcs87, 13\fdg0	& 41	\\
FU Tau 		& 2018-05-20	& 44	& 15.0--313.7  & 2215	& J0426$+$2327, J0423$-$0120, J0423$-$0120 &	1\farcs19$\times$0\farcs92, 6\fdg1	& 39
\enddata
\tablenotetext{\dagger}{On-source integration time.}
\end{deluxetable*}

Analogous to circumstellar disks, circumplanetary disks regulate mass accretion and angular momentum transport as giant planets form, thereby determining the formation timescale, temperature, luminosity, and terminal spin velocity of giant planets (e.g., \citealt{E15,Z15,SM17, Batygin18,Bryan18}). Characterizing circumplanetary disks helps reveal the physical mechanisms involved in giant planet formation as well as the initial conditions and bulk composition of exomoons, providing insight into similar processes that occurred for giant planets in the early Solar System (e.g., \citealt{SH15,Szulagyi17,DS18,Szulagyi18b}). Circumplanetary disks inside gapped protoplanetary disks are expected to be detectable with ALMA across a range of millimeter and sub-millimeter wavelengths (e.g., \citealt{Z16,Z18,Szulagyi18a}), and recently the first such tentative detections have been reported around PDS 70 \citep{Isella19}.

Disks around longer-period planetary-mass companions (PMCs) are promising targets to study the structure and evolution of circumplanetary disks. Discovered in direct imaging surveys, long-period PMCs have masses of $\lesssim$20~$M_{\rm Jup}$ and very wide orbits of $\gtrsim$100 AU. The formation of PMCs remains poorly constrained. While binary-like formation, such as prestellar core collapse \citep{LL76} or disk fragmentation (e.g., \citealt{Kratter10}), is often invoked to explain their existence, PMCs' bottom-heavy mass distribution may imply that they are predominantly formed via core or pebble accretion \citep{Nielsen19,Wagner19}. 

PMC disks are likely to be easier to detect than circumplanetary disks around close-in giant planets because PMCs are widely separated from their hosts and their Hill radii are accordingly larger. Resolved dust and gas observations of PMC disks may reveal substructures that could be linked to exomoons, and allow direct dynamical mass measurements that would enable tests of evolutionary models in the planet-mass regime. Recent photometric, spectroscopic, and polarimetric observations have shown that many young PMCs show signs of active, ongoing accretion (e.g., \citealt{Bowler11,Bowler14,Bowler17,Z14,W15b,W17a,Ginski18,SM18}). However, PMC disks have so far remained undetected in radio observations (e.g., \citealt{Bowler15,M17,R17,W17b,Wolff17,Perez19a}). 

Here we present ALMA 0.88 mm (Band 7) observations of six PMCs (CT Cha b, 1RXS J160929.1--210524 b, ROXs 12 b, ROXs 42B b, DH Tau b, and FU Tau b) and discuss the implications.

\section*{\textbf {\normalsize2. O\lowercase{bservations and} D\lowercase{ata} R\lowercase{eduction}}}
To maximize continuum sensitivity, we performed the survey in Band 7 instead of Band 6 (1.3~mm) as disks emit stronger dust continuum at shorter wavelengths. Our resolution of 0\farcs6--1\farcs2 was sufficient to separate the companions from the primaries, yet not spatially resolve the companion disks (assuming sizes less than the Hill radius) to maximize the sensitivity. Similar strategies have been adopted in previous studies of PMC disks \citep{Bowler15,M17,R17}.

\begin{figure*}[t]
\centering
\figurenum{1}
\includegraphics[width=\linewidth]{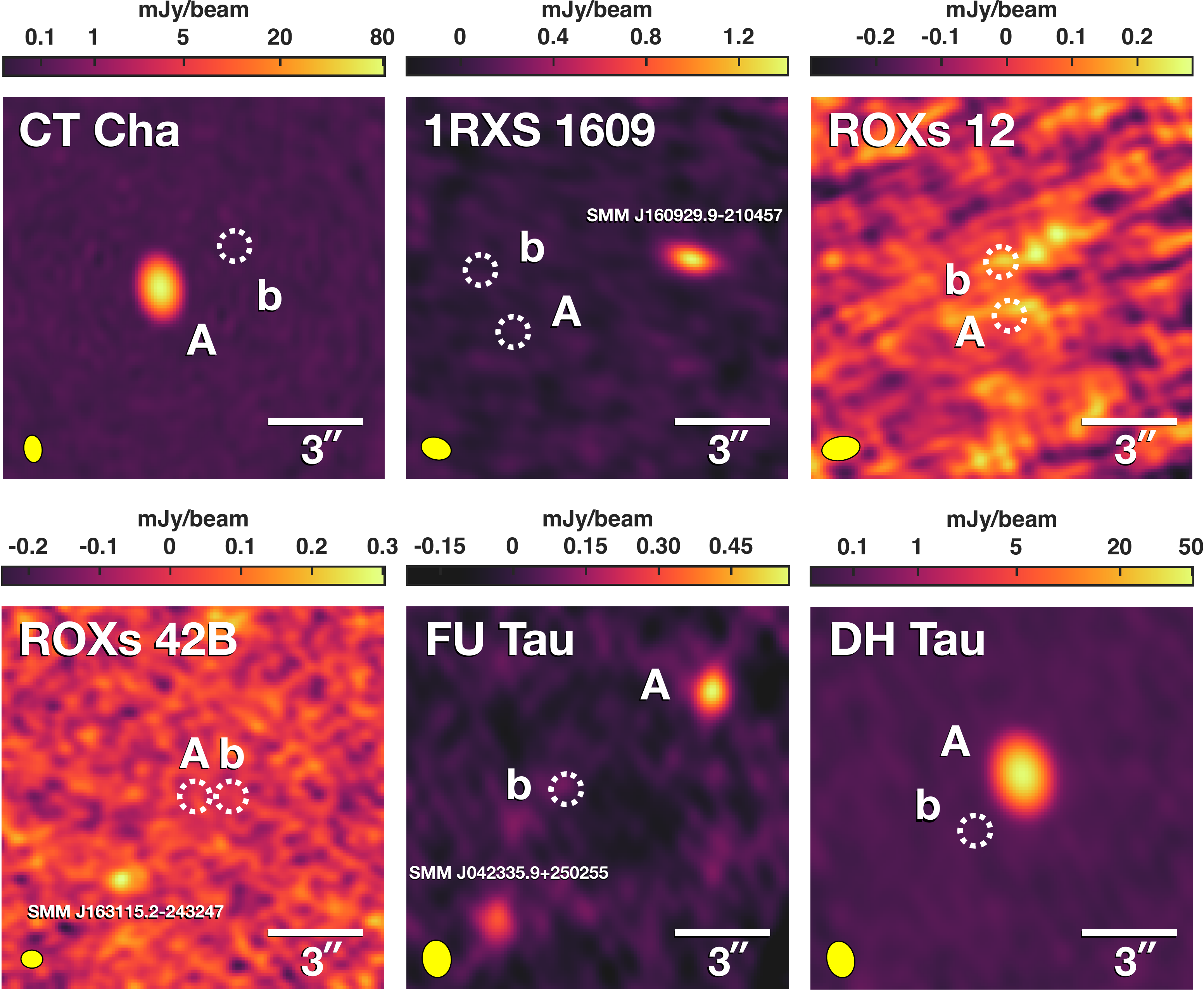}
\caption{ALMA Band 7 images of our six targets. No convincing detection of a PMC disk is seen with our 0.88 mm continuum imaging. The host stars and PMCs are labeled as A and b. Unknown sources, which could be background galaxies, are labeled by their coordinates. CT Cha and DH Tau are shown in a logarithmic scale. North is up and east is to the left.}
\label{fig:source}
\end{figure*}

We observed these systems with three wideband 1.875 GHz windows centered at 333.80, 335.75, and 347.75 GHz to image dust continuum, and one 0.938 GHz window centered at 345.796 GHz, with a resolution of 0.488 MHz (0.423 \kms; Hanning smoothed) to search for ${}^{12}$CO (3--2) emission (ALMA Cycle 4 program \#2016.1.01018.S; PI: B. Bowler). The phase centers for all the observations except FU Tau are at the J2000.0 positions of the primary stars. For FU Tau, the phase center corresponds to the J2000.0 position of the companion because the binary separation is 5\farcs7. We retrieved raw data sets from the ALMA archive and executed the calibration scripts with CASA \citep{McMullin07} versions 4.7.0 (for the data of CT Cha and 1RXS 1609), 4.7.2 (for ROXs 12 and ROXs 42B), and 5.1.1 (for DH Tau and FU Tau) to generate calibrated visibilities. We then applied phase and amplitude self-calibrations to the visibilities of the CT Cha and DH Tau since they had bright enough emission to improve upon the pipeline reduction. We first flagged CO emission and then employed five rounds of phase calibration, starting from a solution interval of 400 s and gradually decreasing to 30 s in subsequent iterations. We then carried out one round of amplitude calibration with a solution interval of 30 s. Self-calibration reduced the background rms by a factor of $\sim$9 for CT Cha and by a factor of $\sim$6 for DH Tau, while the flux densities of both stars only changed by $\lesssim$5\%. Finally, we applied these phase and amplitude solutions to the original data sets in which CO was not flagged. We next used the CASA routine \texttt{tclean} with the multi-frequency synthesis mode and natural weighting to create the primary-beam-corrected continuum maps shown in Figure \ref{fig:source}. For undetected sources (some primary stars and companions), the fluxes and rms uncertainties in Table \ref{tb:RADEC} were first measured in the non-primary-beam-corrected images using apertures of the same areas as the beams, and then the values were divided by the primary beam response at the positions of the sources. We detect the disks of CT Cha A and DH Tau A in CO (3--2). A joint analysis with the CO (2--1) data in \cite{Sheehan19} will be presented in a forthcoming paper. Table \ref{tb:obs} summarizes the observations.

\begin{deluxetable*}{@{}lrlccccrlll@{}}
\tablewidth{\linewidth}
\tablecaption{Properties of Planetary-mass Companions \label{tb:PMCs}}
\tablehead{
\colhead{PMC} &
\colhead{Mass} &
\colhead{Age} &
\colhead{log($L/L_\sun$)} &
\colhead{$T_{\rm eff}$} &
\colhead{$D$} &
\colhead{$\rho$} &
\colhead{PA} &
\colhead{Region} &
\colhead{Disk Markers} &
\colhead{References}
	\\
	\colhead{ } &
	\colhead{($M_{\rm Jup}$)}  &	
	\colhead{(Myr)}  &
	\colhead{ } &
	\colhead{(K)} &	
	\colhead{(pc)} &
	\colhead{(\arcsec)} &
	\colhead{(\degr)} &
	\colhead{ } &
	\colhead{ } &
	\colhead{ }
}
\startdata 
CT Cha b		& $17\pm5$	& 1--3	& $-2.62\pm0.15$ & $2500\pm100$	& $190.72^{+0.78}_{-0.77}$ & 2.68 & 300.0& Cha I	&Pa$\beta$, high $A_V$ 	&	1, 2, 3, 4	\\
1RXS 1609 b	& $12\pm2$	& 8--14	& $-3.38\pm0.06$ & $2000\pm100$	& $139.14^{+1.33}_{-1.31}$ & 2.22 & 27.7	& Upper Sco	&high $A_V$, red $K'$--$L'$   &	2, 3, 5, 6, 7	\\ 
ROXs 12 b 	& $18\pm3$	& 4--10	& $-2.77\pm0.08$ & $3100^{+400}_{-500}$ 	& $136.65^{+0.79}_{-0.78}$ & 1.78 & 9.1& Upper Sco	&red $K'$--$L'$		&	2, 3, 7, 8	\\  
ROXs 42B b	& $9\pm2$	& 1--3	& $-3.10\pm0.06$ & $\sim$1900 	& $143.59^{+1.54}_{-1.51}$ & 1.17 & 270.0& $\rho$ Oph 	&$\cdots$&	2, 3, 7, 9	\\   
DH Tau b		& $14\pm3$	& 1--3 	& $-2.73\pm0.07$ & $2400\pm100$ 	& $134.85^{+1.28}_{-1.26}$ & 2.35 & 139.4& Taurus	&H$\alpha$, Pa$\beta$	&	2, 3, 7, 10	\\  
FU Tau b 		& $20\pm4$	& 1--3 	& $-2.48\pm0.05$ & $\sim$2400 	& $131.20^{+2.65}_{-2.55}$ & 5.69 & 122.8& Taurus	&H$\alpha$, infrared excess	&	2, 3, 11, 12
\enddata
\tablerefs{(1) \cite{W15a}, (2) \cite{BJ18}, (3) This work, (4) \cite{S08}, (5) \cite{W15b}, (6) \cite{PM12}, (7) \cite{K14}, (8) \cite{Bowler17}, (9) \cite{Daemgen17}, (10) \cite{Bonnefoy14}, (11) \cite{Luhman09}, (12) \cite{Todorov14}.}
\end{deluxetable*}

\section*{\textbf {\normalsize3. T\lowercase{argets}}}
\label{sect:PMCs}
In this section we briefly describe each system. To reduce systematics between literature values and incorporate the precise $Gaia$ DR2 distances into the mass estimates, we calculated the companion mass by applying the bolometric correction in \cite{Filippazzo15} to the dereddened $K$-band absolute magnitude. Then the estimated bolometric luminosity was compared with the interpolated age-luminosity-mass grids from the models of \cite{Baraffe15} in a Monte Carlo fashion. Table \ref{tb:PMCs} lists the physical parameters and evidence of disks or mass accretion of these PMCs.

CT Cha is a late-K star in the 1--3 Myr Chamaeleon I star-forming region \citep{HM73,LM13}. Its accretion disk was recently resolved in CO (2--1), enabling a dynamical mass measurement of $0.796^{+0.015}_{-0.014}~M_\sun$~\citep{Sheehan19}. The $17\pm5~M_{\rm Jup}$ companion CT Cha b at a separation of 2\farcs68 (projected separation $\sim$511 AU) was discovered by \cite{S08}, and was found to harbor a disk based on the presence of Pa$\beta$ and Br$\gamma$ lines as well as significant dust extinction \citep{S08,L15,W15a}, but the disk was undetected with ALMA at 1.3 mm \citep{W17b}.

1RXS 1609 is a member of the Upper Scorpius star-forming region \citep{Preibisch98}. The $12\pm2~M_{\rm Jup}$ companion 1RXS 1609 b at a separation of 2\farcs22 ($\sim$309 AU) was discovered by \cite{L08}. It may harbor a disk as inferred from the unresolved {\it Spitzer} 24~\micron~emission \citep{B13}, redder $K'-L'$ than young field dwarfs \citep{K14}, and some optical extinction \citep{W15b}. However, the disk was also not detected with ALMA at 1.3 mm \citep{W17b}.

ROXs 12 lies between the Ophiuchus and Upper Scorpius star-forming regions, with an estimated age of $\sim$6 Myr \citep{Bowler17}. The companion ROXs 12 b at a separation of 1\farcs78 ($\sim$243 AU) was first identified as a candidate companion by \cite{Ratzka05}, and its common proper motion was later confirmed by \cite{K14}. The red $K'-L'$ color of ROXs 12 b may result from a disk \citep{K14}. \cite{Bryan16} found that the orbital motion of ROXs 12 b is consistent with low to moderate eccentricity. \cite{Bowler17} recently identified a tertiary companion with a mass of $\sim$0.5~$M_\sun$ at a separation of $\sim$5100 AU. They also found that the spin axes of A and the tertiary, as well as the orbital axis of b, are likely misaligned. \cite{Sheehan19} recently observed the tertiary with ALMA and derived a disk-based dynamical mass in good agreement with model-dependent mass estimates.

ROXs 42B is a likely member of the $\rho$ Ophiuchus star-forming region \citep{BA92}. The central stars ROXs 42B AB is a close binary separated by a few AU \citep{Simon95,Ratzka05}. The substellar companion ROXs 42B b at 1\farcs17 ($\sim$168 AU) from the binary was identified by \cite{Ratzka05} and confirmed by \cite{K14}. Its orbital motion was also detected by \cite{Bryan16}. Recently, \cite{Bryan18} found that the spin rate of ROXs 42B b is well below the break-up velocity, possibly due to strong magnetic braking from the undetected accretion disks. The 3--5 \micron~photometry of the companion, however, did not reveal significant thermal excess from a disk \citep{Daemgen17}. 

DH Tau is an early-M star in the Taurus star-forming region \citep{Herbig77}. The $14\pm3~M_{\rm Jup}$ companion DH Tau b at a separation of 2\farcs35 ($\sim$317 AU), discovered by \cite{Itoh05}, has H$\alpha$, Pa$\beta$, and optical continuum excess indicative of an accretion rate of $\sim$$10^{-12}~M_\sun$~yr$^{-1}$ \citep{Bonnefoy14,Z14}. However, the companion disk was not detected with either ALMA and NOEMA at 1.3 mm \citep{W17b,Wolff17}. 
 
FU Tau is a young binary system likely in front of the Barnard 215 dark cloud \citep{Barnard27}. The primary FU Tau A could be a low-mass star or a high-mass brown dwarf as its mass was estimated as $\sim$80~$M_{\rm Jup}$ \citep{Stelzer13}. Multi-wavelength tracers ranging from X-ray, optical spectroscopy, to 3 cm free-free emission have indicated that FU Tau A is actively accreting and may also have outflows \citep{Stelzer10,Stelzer13,Rodriguez17}. The companion FU Tau b, 5\farcs69 ($\sim$747 AU) southeast of FU Tau A, was discovered by \cite{Luhman09}. It has strong infrared excess indicating the presence of a circum-substellar disk \citep{Luhman09}. With the revised $K$-band magnitude in \cite{Luhman10}, we estimate a mass of $20\pm4~M_{\rm Jup}$ for FU Tau b.

\begin{deluxetable}{@{}l@{\hspace{6pt}}l@{\hspace{6pt}}l@{\hspace{6pt}}r@{ $\pm$ }l@{}}
\tablecaption{Coordinates and 0.88 mm Flux Densities of the Host Stars, PMCs, and Unknown Sources \label{tb:RADEC}}
\tablehead{
\colhead{Object} &
\colhead{$\alpha$} &  
\colhead{$\delta$} &
\multicolumn2r{$F_\nu$ (mJy)} 
}
\startdata 
CT Cha 				& 	$11^{\mathrm{h}}$04$^{\mathrm{m}}$09\fs00	&  $-$76\degr27\arcmin19\farcs3 & 		94.0    &  9.4   		\\  
CT Cha b				& 	$11^{\mathrm{h}}$04$^{\mathrm{m}}$08\fs34	&  $-$76\degr27\arcmin18\farcs0 &		0.019  &  0.052         \\
1RXS 1609 			& 	$16^{\mathrm{h}}$09$^{\mathrm{m}}$30\fs30	&  $-$21\degr04\arcmin59\farcs3 &		0.007  &  0.041		\\  
1RXS 1609 b			& 	$16^{\mathrm{h}}$09$^{\mathrm{m}}$30\fs37	&  $-$21\degr04\arcmin57\farcs4 &		0.011  &  0.045		\\  
ROXs 12 				& 	$16^{\mathrm{h}}$26$^{\mathrm{m}}$28\fs03	&  $-$25\degr26\arcmin48\farcs2 &		0.078  &  0.070 	\\  
ROXs 12 b			& 	$16^{\mathrm{h}}$26$^{\mathrm{m}}$28\fs05	&  $-$25\degr26\arcmin46\farcs4 &		0.098  &  0.071		\\  
ROXs 42B			& 	$16^{\mathrm{h}}$31$^{\mathrm{m}}$15\fs01	&  $-$24\degr32\arcmin44\farcs1 & 		$-$0.001 &  0.043  	\\  
ROXs 42B b			& 	$16^{\mathrm{h}}$31$^{\mathrm{m}}$14\fs93	&  $-$24\degr32\arcmin44\farcs1 &		$-$0.002 &  0.043	\\  
DH Tau 				& 	$04^{\mathrm{h}}$29$^{\mathrm{m}}$41\fs56	&  $+$26\degr32\arcmin57\farcs7 &		52.3	  &   5.2		\\  
DH Tau b				& 	$04^{\mathrm{h}}$29$^{\mathrm{m}}$41\fs68	&  $+$26\degr32\arcmin56\farcs0 & 		0.017  &  0.041		\\  
FU Tau  				& 	$04^{\mathrm{h}}$23$^{\mathrm{m}}$35\fs40	&  $+$25\degr03\arcmin02\farcs4 & 		0.53	  &   0.09		\\  
FU Tau b				& 	$04^{\mathrm{h}}$23$^{\mathrm{m}}$35\fs75	&  $+$25\degr02\arcmin59\farcs2 &		0.033 &   0.039		\\
SMM~J042335.9$+$250255	& $04^{\mathrm{h}}$23$^{\mathrm{m}}$35\fs91	&  $+$25\degr02\arcmin55\farcs2 &		0.40   &   0.09		\\    
SMM~J160929.9$-$210457 	& $16^{\mathrm{h}}$09$^{\mathrm{m}}$29\fs89	&  $-$21\degr04\arcmin57\farcs0 &		1.86   &   0.21		\\
SMM~J163115.2$-$243247	& $16^{\mathrm{h}}$31$^{\mathrm{m}}$15\fs18	&  $-$24\degr32\arcmin46\farcs7 &		0.42   &   0.10		\\	 	    
\enddata
\tablecomments{The coordinates are derived for the epochs of ALMA observations (Table \ref{tb:obs}). Flux uncertainties of the detected sources include a 10\%~absolute flux calibration error.}
\end{deluxetable}

\begin{figure}[h]
\centering
\figurenum{2}
\includegraphics[width=\columnwidth]{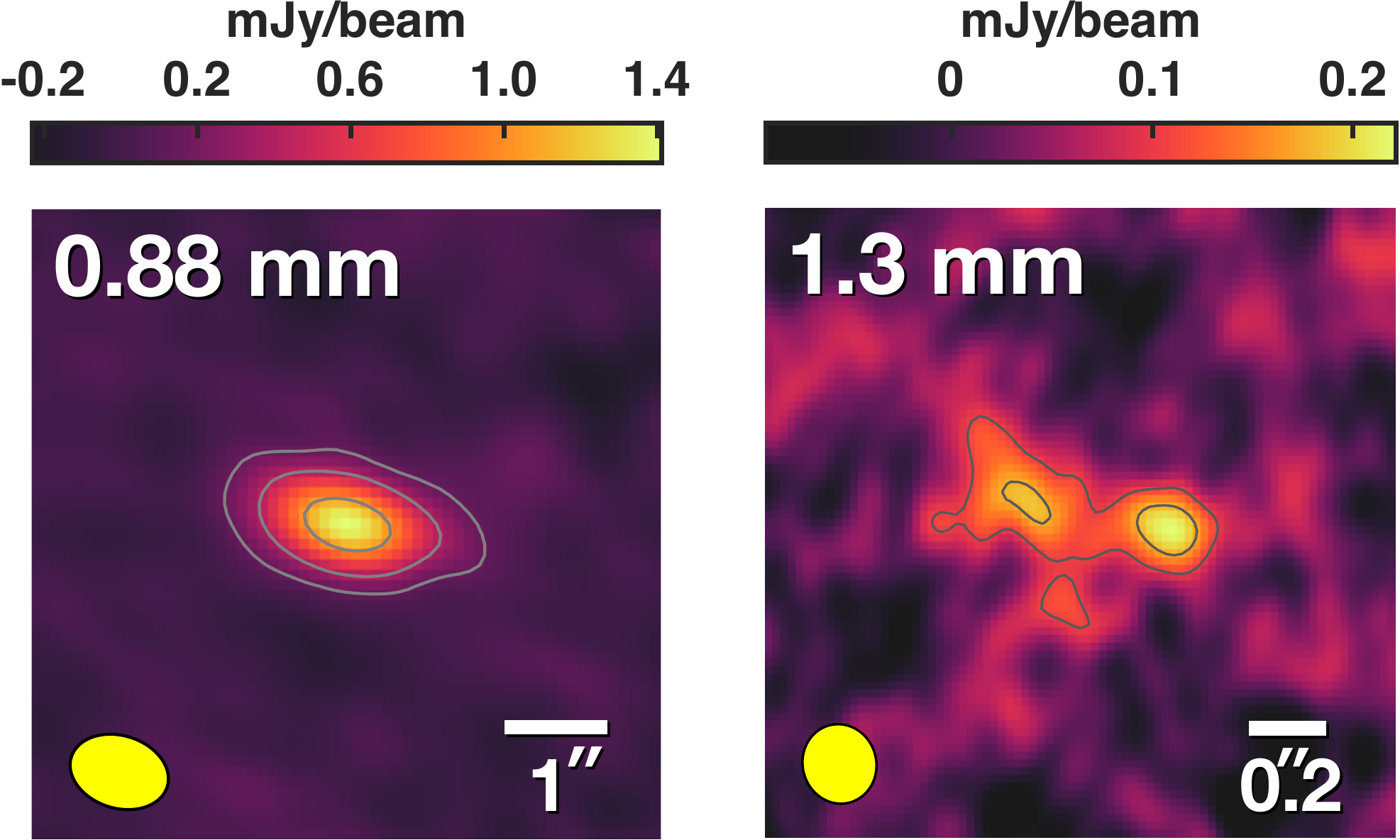}
\caption{Continuum observations of the bright source $\sim$6\arcsec~west from 1RXS 1609 at 0.88 mm (left) and 1.3 mm (right; data from \citealt{W17b}). The source appears spatially resolved into two components at 1.3 mm, which we speculate may be two interacting galaxies or two star-forming regions in a galaxy. Contours are [3, 8, 20] $\times$~51~$\mu$Jy beam$^{-1}$ in the left panel, and [3, 5] $\times$~35~$\mu$Jy beam$^{-1}$ in the right panel.}
\label{fig:gal}
\end{figure}

\begin{figure*}[t]
\centering
\figurenum{3}
\includegraphics[width=\linewidth]{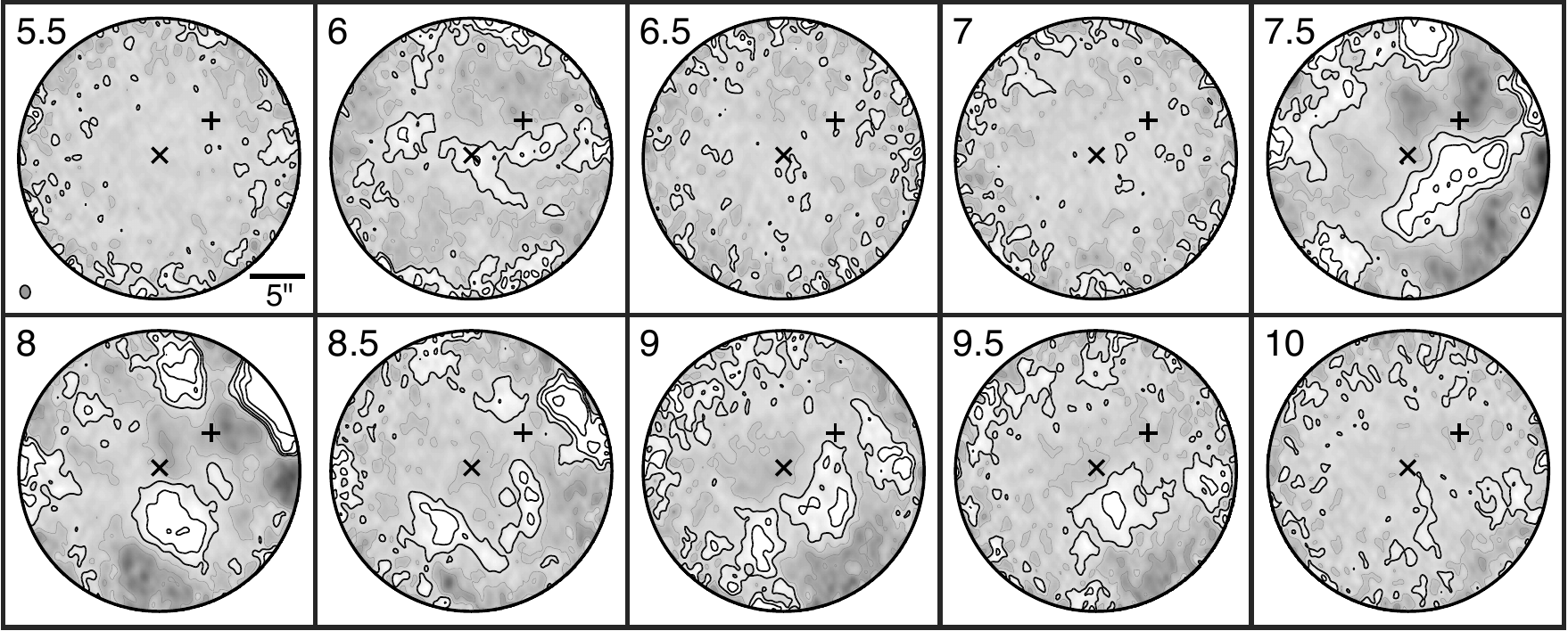}
\caption{The distribution of CO emission around FU Tau is more consistent with diffuse emission from the Barnard 215 dark cloud rather than a molecular outflow from FU Tau A, as suggested by previous low-resolution observations. The radial velocity is shown at the upper left corner of each channel. The positions of FU Tau A and b are marked with $+$ and $\times$, respectively. The synthesized beam ($1\farcs17\times0\farcs91$, PA$=6\fdg3$) is plotted at the bottom left corner of the first channel. Contours are plotted in $-10$, $-3$, 3, 10, 20, and 40$\sigma$, with 1 $\sigma = $ 2.3 mJy beam$^{-1}$ measured at the image center. The maximum recoverable scale is calculated as $0.6~\lambda/L_{\rm min~baseline}\sim7\farcs3$. The edge of each image corresponds to a primary beam gain level of 0.2. North is up and east is left.}
\label{fig:channel}
\end{figure*}

\section*{\textbf {\normalsize4. R\lowercase{esults}}}
\subsection*{4.1. C\lowercase{ontinuum} E\lowercase{mission:} PMC\lowercase{s} \lowercase{and} H\lowercase{ost} S\lowercase{tars}}
As shown in Figure \ref{fig:source}, we do not convincingly detect any PMC disks in 0.88 mm dust continuum (dashed circles labeled ``b''). Table \ref{tb:RADEC} lists the measured flux levels of PMCs. The 3$\sigma$ upper limits for the 0.88 mm flux density span 120--210 $\mu$Jy assuming they are point sources. On the other hand, disks around the three host stars CT Cha A, DH Tau A, and FU Tau A are clearly seen. To our knowledge, this is the first millimeter detection of FU Tau A.

We use the software package {\tt pdspy} \citep{Sheehan19} to fit an elliptical gaussian to the visibilities and derive 0.88 mm continuum fluxes of $94.0\pm9.4$ mJy, $52.3\pm5.2$ mJy, and $0.53\pm0.09$ mJy for CT Cha A, DH Tau A, and FU Tau A, respectively. Similar values were also obtained using the CASA task {\tt uvmodelfit} when the primary beam response is accounted for. Flux errors are generally dominated by a 10\%~uncertainty in ALMA absolute flux calibration. Our results are consistent with previous 0.88 mm flux measurements of CT Cha A ($104.78\pm0.60$ mJy; \citealt{Pascucci16}) and DH Tau A ($47\pm4$ mJy; \citealt{Harris12}) at about 1$\sigma$ level. This discrepancy might simply arise from absolute flux calibration; free-free emission from ionized disk material is usually minor at 0.88 mm (e.g., \citealt{Eisner18}).

Table \ref{tb:RADEC} also lists the coordinates of the host stars and PMCs at the time of the ALMA observations as calculated from the $Gaia$ DR2 coordinates, proper motions, position angles, and separations.

\subsection*{4.2. C\lowercase{ontinuum} E\lowercase{mission:} B\lowercase{ackground} O\lowercase{bjects}}
Several continuum sources in our maps seem to have no optical or infrared counterparts. Some of them could be dusty star-forming galaxies at high redshift, similar to the object spatially coincident with the debris disk of HD 95086 \citep{Su17}. The object SMM J042335.9$+$250255 at $\alpha=04^{\mathrm{h}}$23$^{\mathrm{m}}$35\fs91 and $\delta=+$25\degr02\arcmin55\farcs2, $\sim$10\arcsec~southeast of FU Tau A, has a 0.88 mm flux of $0.40\pm0.09$ mJy. The source SMM J163115.2$-$243247 at $\alpha=16^{\mathrm{h}}$31$^{\mathrm{m}}$15\fs18 and $\delta=-$24\degr32\arcmin46\farcs7, 3\farcs5~southeast of ROXs 42B, has a flux of $0.42\pm0.10$ mJy. It shares a similar PA of about 138\degr~to the background near-infrared source identified by \cite{Bryan16} but with a wider separation from ROXs 42B (3\farcs5 versus 3\arcsec; see their Figure 1), an offset that cannot be accounted for by the star's proper motion ($\mu_\alpha~{\rm cos}~\delta=-6.5\pm0.2$ mas yr$^{-1}$ and $\mu_\delta=-23.2\pm0.1$ mas yr$^{-1}$; \citealt{GaiaDR2}). 

The bright source SMM J160929.9$-$210457 at $\alpha=16^{\mathrm{h}}$09$^{\mathrm{m}}$29\fs89 and $\delta=-$21\degr04\arcmin57\farcs0, $\sim$6\arcsec~northwest of 1RXS 1609, was also detected in the 1.3 mm survey of \cite{W17b}. It has a flux density of $1.86\pm0.21$ mJy at 0.88 mm and $0.68\pm 0.17$ mJy at 1.3 mm. The 1.3 mm flux density was derived using aperture photometry due to the irregular morphology, which made it difficult to reliably measure flux density using an elliptical gaussian fit. Interestingly, it appears resolved into west and east components at 1.3 mm (Figure \ref{fig:gal}), which could represent two star-forming regions within a galaxy, or perhaps a pair of merging galaxies. Indeed, given flux-dependent number densities of such galaxies at 0.88 mm \citep{C14}, $\sim$2.5/$\rm{arcmin}^2$ at 1.9 mJy and $\sim$25/$\rm{arcmin}^2$ at 0.4 mJy, we would expect to detect $\sim$1 and $\sim$10 galaxies, respectively, for six ALMA pointings (assuming an 18\arcsec~primary beam).

\subsection*{4.3. CO E\lowercase{mission:} O\lowercase{utflow from} FU T\lowercase{au} A?}
Figure \ref{fig:channel} shows the channel maps of the FU Tau system in the LSRK velocity frame. CO emission is seen from 6--10~\kms; however, it does not spatially coincide with either A or b. This suggests that it is probably not associated with both components, but is instead diffuse emission from the surrounding Barnard 215 dark cloud \citep{Barnard27}. Although FU Tau A has active accretion, we do not find concrete evidence of gas emission from the disk in our data. The CO (2--1) emission detected by the IRAM-30m telescope \citep{Monin13} is probably from this cloud emission as well, rather than a molecular outflow from FU Tau A.

\begin{figure*}[t]
\centering
\figurenum{4}
\includegraphics[width=\linewidth]{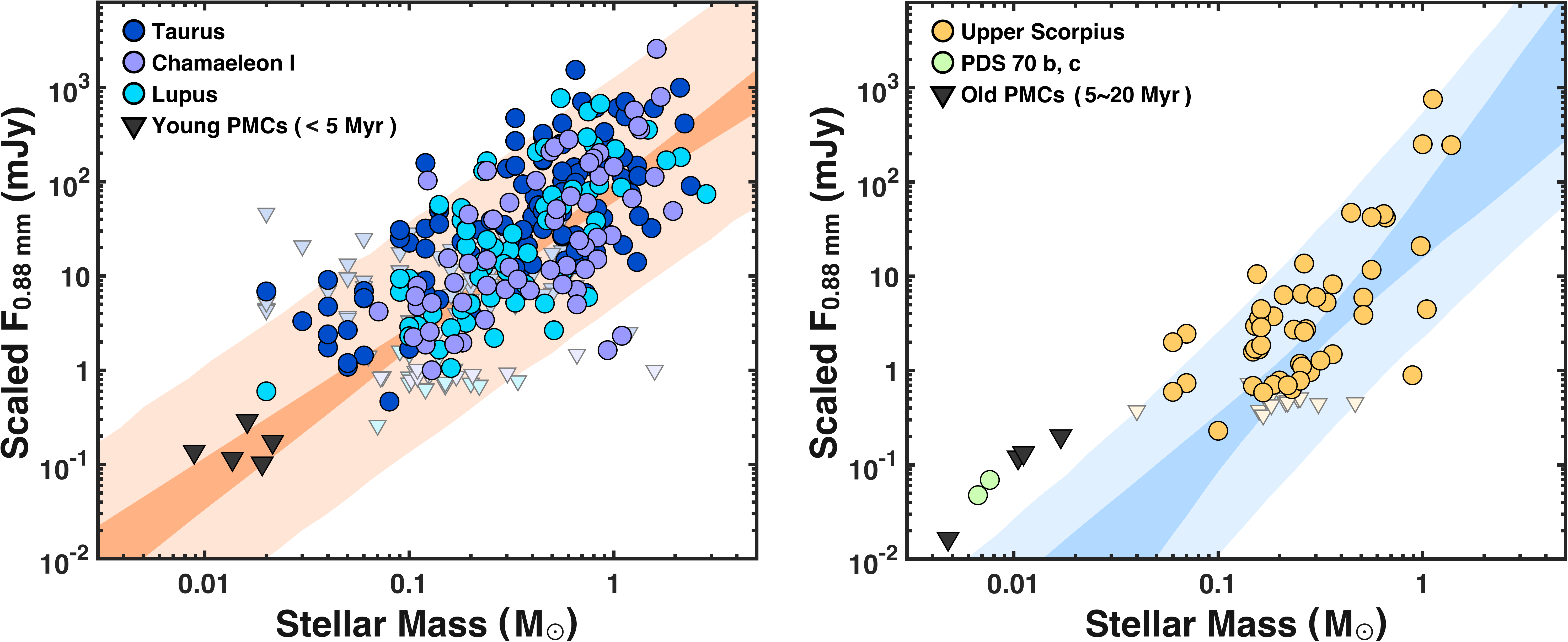}
\caption{Disk fluxes as a function of host mass at 0.88 mm. All fluxes and 3$\sigma$ upper limits (triangles) are scaled to a common distance of 140 pc. We perform Bayesian linear regression following \cite{Kelly07} to the stellar and brown dwarf disks (including non-detections). The dark shaded regions are the 95\% confidence intervals for the best-fit relations, and the light shaded regions mark the same relations but adding the intrinsic scatter terms. In the left panel we compare young companions (FU Tau b, DH Tau b, CT Cha b, ROXs 42B b, GQ Lup b) with the $F_{\rm disk}$--$M_\star$ relation of young star-forming regions (Taurus, Chamaeleon I, Lupus). In the right panel we compare older companions (2M1207 b, ROXs 12 b, 1RXS 1609 b, GSC 6214-210 b), as well as the the tentative ALMA detections of PDS 70 b, c \citep{Isella19}, with the relation of Upper Scorpius. Our observations tentatively suggest that young PMC disks are systematically fainter, as no disks are seen between the mean relation and the upper boundary of the intrinsic scatter. For more evolved systems, our PMC limits are not sensitive enough to draw any meaningful inferences. However, the disks around the PDS 70 giant planets appear overluminous than the mean relationship.}
\label{fig:LmmMs}
\end{figure*}


\section*{\textbf {\normalsize5. D\lowercase{iscussion}}}
\subsection*{5.1. D\lowercase{isk} F\lowercase{lux}-H\lowercase{ost} M\lowercase{ass} R\lowercase{elationship}} 

The dominant physical processes regulating the coevolution of planets and their environments may leave observable imprints on the bulk properties of disks. Disk-host scaling relationships, along with their age and stellar-mass dependences, therefore have important implications for the observed exoplanet population (e.g., \citealt{A13,Ansdell17,Pascucci16,Barenfeld16,Andrews18}). The behaviors of scaling relationships in the planet-mass regime provide clues about the physical properties of circumplanetary disks as well as the formation of satellites. Any deviations from the stellar relationships may imply that circumplanetary disks have different mass distributions, sizes, substructures, or evolutionary timescales. It may also imply that the coevolution of satellites and circumplanetary disks are not self-similar to that of planets and protoplanetary disks.

Here we investigate the behavior of the disk flux-host mass ($F_{\rm disk}$--$M_\star$) relation toward the planetary-mass regime. We opt not to convert continuum emission to dust mass, as the observed millimeter spectral index of $F_{\rm disk}\propto\lambda^{-2.2}$ (e.g., \citealt{Ribas17}) suggests that typical assumptions used to make the conversion (optically thin, opacity $\propto\lambda^{-1}$; e.g., \citealt{B90}) may be incorrect. Recent radiative transfer disk modeling also finds that dust emission is generally optically thick for circum(sub)stellar disks (e.g., \citealt{BE19,Rab19}). In addition, small disks are likely warmer and optically thicker than large disks; as a result, dust masses derived from the optically thin assumption may be underestimated by a factor of 1--5 (e.g., \citealt{BE19}). Finally, \cite{Liu19} and \cite{Zhu19} demonstrated that an optically thick disk may look optically thin due to dust scattering.

Figure \ref{fig:LmmMs} shows the $F_{\rm disk}$--$M_\star$ relationship at 0.88 mm for young ($<$5 Myr; left panel) and older (5--20 Myr; right panel) systems. All the fluxes and upper limits in the figure have been scaled to a common distance of 140 pc via $F_{\rm disk}=F_{\rm disk}\times(\mbox{distance}/140~\mbox{pc})^2$. PMC 3$\sigma$ limits at 0.88 mm are either from this survey or previous ALMA observations ($<$200~$\mu$Jy for GSC 6214-210 b, \citealt{Bowler15}; $<$150~$\mu$Jy for GQ Lup b, \citealt{M17}; $<$78~$\mu$Jy for 2M1207 b, \citealt{R17}). We also include the recent tentative detections of circumplanetary disks around PDS 70 b and c ($73\pm19~\mu$Jy and $106\pm19~\mu$Jy; \citealt{Isella19}), with planet masses adopted from \cite{Keppler18} and \cite{Haffert19}. Disk fluxes of stars and brown dwarfs are compiled from millimeter studies of star-forming regions, including Taurus \citep{A13,Ricci14,WD18}, Chamaeleon I \citep{Pascucci16,Long18}, Lupus \citep{Ansdell16,Sanchis20}, and Upper Scorpius\footnote{For the samples in \cite{Barenfeld16}, only the \textit{primordial} disks were included in our analysis.} \citep{Barenfeld16,V16}. Sources with no {\it Gaia} DR2 distances or that are clearly foreground/background objects are excluded. Stellar masses are adopted from the literature, but dynamical masses derived from the Keplerian rotation of gas are used when available \citep{Sheehan19,Simon19}. 

To explore if PMC disks largely follow the overall trend for protoplanetary disks, we perform Bayesian linear regression to the stellar and brown dwarf disks (including non-detections) using the {\tt linmix} package \citep{Kelly07} and extrapolate the $F_{\rm disk}$--$M_\star$ relationship from stellar to planetary regime. The best-fit relations with and without the Gaussian-distributed intrinsic scatter terms are plotted as the dark and light shaded regions. In the left panel, the slope, intercept, and scatter of log($F_{\rm disk}$) are $1.54^{+0.21}_{-0.21}$, $1.88^{+0.14}_{-0.13}$, and $0.74^{+0.08}_{-0.07}$ dex, respectively, where the uncertainties represent the 95\% confidence intervals. We note that as shown in the Appendix and Figure \ref{fig:LmmMs_SFR}, Taurus disks seem to have a shallower $F_{\rm disk}$--$M_\star$ relation than that of the Chamaeleon I and Lupus disks. However, it is not clear whether this is of astrophysical origin or simply because few disks with $M_\star<0.1~M_\sun$ have been detected in Chamaeleon I and Lupus. The inhomogeneity of stellar mass determinations could also potentially affect our linear regression analysis (e.g., \citealt{WD18}), but we do not expect this to have a strong impact on our results. For the Upper Sco disks in the right panel, the slope, intercept, and scatter are $2.37^{+0.73}_{-0.71}$, $1.58^{+0.44}_{-0.43}$, and $0.67^{+0.19}_{-0.13}$ dex, consistent with \cite{Ansdell17}.

Figure \ref{fig:LmmMs} shows that young PMCs have distance-scaled 3$\sigma$ limits of 100--300~$\mu$Jy at 0.88 mm. Given that the majority of these PMCs have evidence of disks or mass accretion, yet appear absent between the best-fit relation and the upper envelope in the stellar regime (represented by the 95\% confidence interval of the intrinsic scatter), it remains a possibility that the (presumably) brightest PMC disks are in fact underluminous and may have different bulk properties from their stellar counterparts. More sensitive observations are needed to test this hypothesis. Alternatively, as individual disks in young binary systems of $\lesssim$300 AU separations have on average $\sim$5 times lower millimeter fluxes than that of single stars (e.g., \citealt{Harris12,Akeson19}), young PMC disks may instead be in agreement with the $F_{\rm disk}$--$M_\star$ relationship of binary stars. Deep ALMA continuum imaging with rms $\lesssim$10~$\mu$Jy ($\gtrsim$5 hr on-source at 0.88 mm) will be needed to definitively test whether young PMC disks indeed deviate from the scaling relation of single stars.

In contrast, \cite{Barenfeld19} recently showed that in the more evolved Upper Scorpius star-forming region (age $\sim$10 Myr; \citealt{PM12}), the millimeter fluxes of disks around single and binary stars are statistically indistinguishable. This implies that binarity unlikely plays an important role in subsequent disk evolution, and single and binary stars may follow the same $F_{\rm disk}$--$M_\star$ relationship at ages of $>$5 Myr. Current ALMA observations, however, are not able to determine if older PMC disks are discrepant from the Upper Sco correlation (right panel of Figure \ref{fig:LmmMs}). A sensitivity of $\lesssim 1 \mu$Jy will be required, but this will be very challenging to achieve given the current capabilities of ALMA.

\subsection*{5.2. I\lowercase{mplications}} 
Non-detections from this survey as well as other programs have demonstrated that PMC disks appear faint in dust continuum emission. As the radial drift of dust grains around substellar objects can be much faster than around stars (e.g., \citealt{Pinilla13,Z18}), the dust disks of these companions are likely very compact and optically thick (e.g., \citealt{W17b,Rab19}). As optically thick emission scales with disk area, our 0.88 mm and 1.3 mm surveys imply that these dust disks have radii $<$0.5 AU (cf. \citealt{W17b}). It is also possible that dust grains in PMC disks have been highly depleted due to this rapid inward drift. This could potentially hinder satellite formation unless there is a dust trap in the disk to form satellite seeds (e.g., \citealt{DS18}).

While a compact and/or dust-depleted disk is faint in radio continuum, \cite{Rab19} find that the gas disk may still extend to 1/3 Hill radius, which is the characteristic scale of companion disks under dynamical interaction (e.g., \citealt{AB09}) and is typically $\gtrsim$10 AU for wide-orbit PMCs at hundreds of AU from their hosts. The synthetic observations in \cite{Rab19} further show that such a large gas-rich disk can be detected by ALMA with $\gtrsim$5 hr on-source time in Bands 6 and 7. PMC disks may therefore exhibit both extremely high gas-to-dust {\it mass} and {\it size} ratios due to a combination of tidal truncation, fast radial drift, and/or a lack of dust traps. As a comparison, for the Lupus circumstellar disks the mass ratio is mostly 1--10 \citep{Miotello17}, and the size ratio is $\sim$2 on average \citep{Ansdell18}.

Interestingly, free-floating planetary-mass objects, which might have formed like single stars, may be able to retain larger and brighter dust disks. For instance, the isolated 13~$M_{\rm Jup}$~object OTS 44 has a 1.3 mm flux of $\sim$100 $\mu$Jy \citep{B17}, which translates to $\sim$230 $\mu$Jy at 0.88 mm with a spectral index of 2.2. This implies that radial drift timescale is intrinsically longer than that of PMC disks, or there are disk substructures to trap dust grains at a wider radius. Future sensitive ALMA observations can examine if free-floating planets, as a population, also follow the stellar $F_{\rm disk}$--$M_\star$ relationship.

\acknowledgements
We thank the anonymous referee for constructive suggestions. Y.-L.W. is grateful to the 51 Pegasi b postdoctoral fellowship from the Heising-Simons Foundation. B.P.B. acknowledges support from the National Science Foundation grant AST-1909209. Y.-L.W. also thanks Brandon Kelly, Kimberly Ward-Duong, and Yueh Wang for their help in the linear regression analysis. This paper makes use of the following ALMA data: ADS/JAO.ALMA \#2013.1.00487.S, \#2015.1.00773.S, and \#2016.1.01018.S. ALMA is a partnership of ESO (representing its member states), NSF (USA) and NINS (Japan), together with NRC (Canada), NSC and ASIAA (Taiwan), and KASI (Republic of Korea), in cooperation with the Republic of Chile. The Joint ALMA Observatory is operated by ESO, AUI/NRAO, and NAOJ. The National Radio Astronomy Observatory is a facility of the National Science Foundation operated under cooperative agreement by Associated Universities, Inc.

\facilities{ALMA}
\software{CASA \citep{McMullin07}, {\tt pdspy} \citep{Sheehan19}}, {\tt linmix} \citep{Kelly07}\\ \\ \\ \\ \\ \\ \\

\begin{figure*}[t]
\centering
\figurenum{5}
\includegraphics[width=\linewidth]{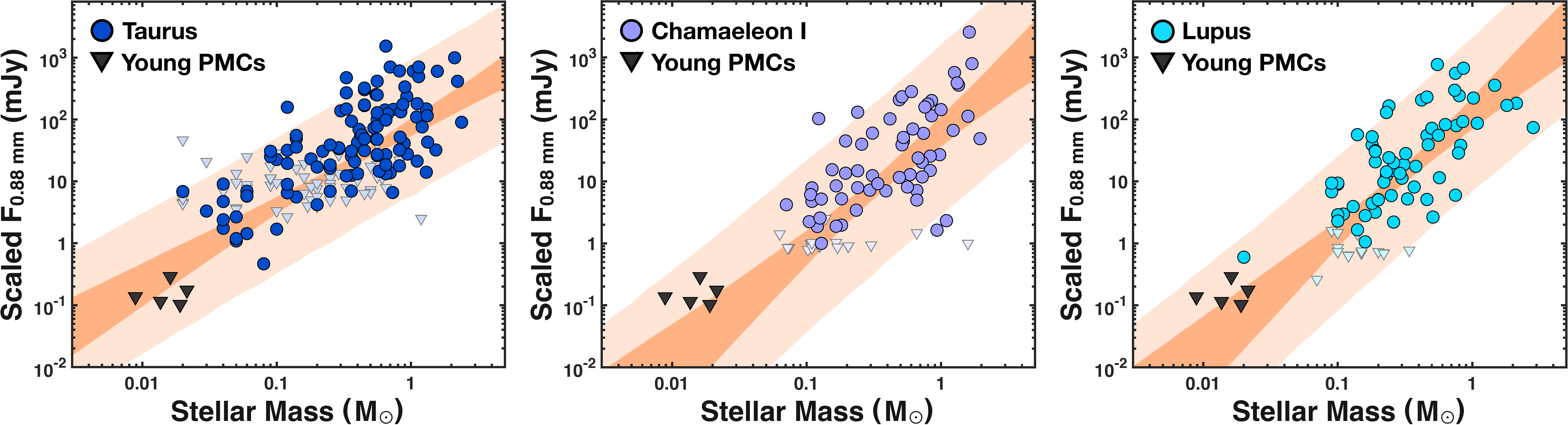}
\caption{The $F_{\rm disk}$--$M_\star$ relationship for individual young star-forming regions. The black triangles are PMC upper limits. The dark shaded regions mark the 95\% confidence intervals of the best-fit relations, and the light shaded regions show the same relations but including the additional scatter terms from the linear regression analysis. For Taurus, the best-fit values and the 95\% confidence intervals for the slope, intercept, and scatter are $1.26^{+0.26}_{-0.24}$, $1.87^{+0.17}_{-0.17}$, and $0.66^{+0.10}_{-0.08}$ dex, respectively. For Chamaeleon I, the values are $1.86^{+0.52}_{-0.49}$, $1.82^{+0.30}_{-0.30}$, and $0.83^{+0.18}_{-0.15}$ dex. For Lupus, the values are $1.91^{+0.47}_{-0.48}$, $2.08^{+0.30}_{-0.31}$, and $0.73^{+0.16}_{-0.12}$ dex.}
\label{fig:LmmMs_SFR}
\end{figure*}
\appendix
The slope of the $F_{\rm disk}$--$M_\star$ relation may be sensitive to the adopted distances and stellar masses, or it may even vary between star-forming regions of similar ages. It is also not clear that if the relation would change in the substellar regime. In Figure \ref{fig:LmmMs_SFR} we show the individual relationships in Taurus, Chamaeleon I, and Lupus. Our fitted slopes are similar to that of the $M_{\rm dust}$--$M_\star$ relation in previous studies (e.g., \citealt{Ansdell17,Long18}). We note that the Taurus slope becomes steeper, from $1.26^{+0.26}_{-0.24}$ to $1.50^{+0.45}_{-0.43}$, when excluding objects with masses less than 0.1~$M_\sun$. Future brown dwarf studies in Chamaeleon I and Lupus can help determine whether the mean $F_{\rm disk}$--$M_\star$ relation flattens out at substellar masses, or if Taurus has an intrinsically distinct relation.

\end{document}